# Providing a Timely Review of Input Demographics to Advisory Committees[1]

Lead Author: Dara Norman NOAO dnorman@noao.edu

Co-Authors: T. J. Brandt, GSFC, t.j.brandt@nasa.gov,
Nancy D. Morrison, University of Toledo, nancy.morrison@utoledo.edu,
Sarah Tuttle, University of Washington, tuttlese@uw.edu,
Julie Rathbun, PSI, rathbun@psi.edu,
Zach Berta-Thompson, University of Colorado, Zachory.BertaThompson@Colorado.edu,
Edmund Bertschinger, MIT, edbert@mit.edu,
Nancy Chanover, NMSU, nchanove@nmsu.edu,
Karen Knierman, ASU, karen.knierman@asu.edu,
Aparna Venkatesan, University of San Francisco, avenkatesan@usfca.edu,
Kim Coble, San Francisco State University, kcoble@sfsu.edu,
Jonathan Fraine, STScI, jfraine@stsci.edu ,
Adam Burgasser, University of California, San Diego, aburgasser@ucsd.edu,
Ivelina Momcheva, STScI, ivelina.momcheva@gmail.com,
Marie Lemoine-Busserolle, Gemini Observatory, mbussero@gemini.edu

Type of Activity: State of the Profession

---

[1] The views expressed represent the personal views of the authors, contributors, and endorsers and not necessarily the views of their affiliated institutions.

**Executive Summary and Recommendations:**
Organizations that support science (astronomy) such as federal agencies, research centers, observatories, academic institutions, societies, etc. employ advisory committees and boards as a mechanism for reviewing their activities and giving advice on practices, policies and future directions.  As with any scientific endeavor, there is concern over complementing these committees with enough members who have as broad a range of expertise and understanding as possible, so that bias is mitigated. However, for a number of reasons (logistical, practical, financial, etc.), committees can also not be infinitely large and thus trade-offs must be made.   It is often recognized that conflicts of interest must be acknowledged within these committees, but what is not often recognized it the potential for unmitigated biases and "group think" that can be introduced as part of these committees.

In this white paper, we ***recommend that advisory committees that collect community input, (e.g., the Decadal Survey review committee), also collect, compile and review input demographic data before finalizing reports, (e.g., the final 2020 Decadal Survey Report). A summary of these data should be released alongside the final survey report.*** This information would enable the committee to understand potential 'blind spots' and biases of the data collection phase and inform future data collections of any barriers that affect the omission of perspectives from various demographics.

**Introduction:**
   The process by which (astronomical) scientific research is assessed and prioritized is done by peer review.    The goals of this process are to ensure that research being undertaken on the public's behalf is based on the scientific merits of the work, that is, it is conducted responsibly, ethically, is reported as accurately as possible and produces (or is likely to produce) results that will advance the field of study.   An additional goal of peer review on advisory panels is to optimize the use of limited resources to promote scientific discovery.   A rigorous application of the process requires the perspectives and knowledge of a range of experts in the field, particularly since prioritization of scientific goals is a critical requirement given funding constraints.

The details of the process of peer review have evolved since its beginnings as standard scientific practice has advanced (e.g., Csiszar, A., 2016*).*  Procedures and policies to confront, for example, conflict of interest and bias irrelevant to merit, etc. are now employed as common best practices.

Scientists are familiar with the concern for bias in research and strive to eliminate and/or mitigate the effects of bias with additional information, approaches, and methods. This white paper puts forth recommendations for doing the same with advisory reviews of priorities in directing scientific resources by providing panels and committees with **timely demographic information about who has given input to the review process, and publishing this information along with committee membership demographics.**

**A More Complete Review:**
Organizations, like the US National Academies and others, that convene panels and commission advisory reports to inform US federal agencies and the US Congress, have a study process in place that is designed to be as independent and objective as possible (for example, see http://www.nationalacademies.org/studyprocess/?_ga=2.189515371.1787766582.1514415258-860307303.1514415258 ) and encompass, "an appropriate range of expertise for the task" as well as a "balance of perspectives" among the committee members, while keeping in mind that committees also cannot be arbitrarily large. Funding agencies like NASA also use such advisory committees and community input in their Federal Advisory Committee Act (FACA) process.[2]

During the study process the committee will gather information. In astronomy and astrophysics, this is often done by soliciting position (white) papers and other communications from the larger scientific community. However, if there are unintended biases in the gathering of information, additional biases can be introduced into the deliberation phase of the study process and thus into the advice given in the formal report. As an example to illustrate the concern, a recent NRC committee charged with discussing the future development of astronomy resources[3] had no committee members from small teaching universities.[4] The main way the committee chose to get input from the community to inform their work was through submitted white papers. However, the only timescale for submission of these positional papers overlapped with the start of the academic school year in early October, when faculty with significant teaching loads were least likely to be able to contribute a response. There was thus a presumably unintended bias created in who might be able to participate in the initial information gathering phase and thus contribute to the overall discussion of the issues and priorities.  Unfortunately, there seems to have been little concern raised within the committee about the effects of input information bias because of this deadline choice and no suggestion for additional broad opportunities for input or alternative deadlines.[5] In this example, identity diversity of the committee members (who they are, i.e., their affiliation) mattered as much as their cognitive diversity (what they know) for their capacity to recognize and remove bias in the collection of data about how the full community's resource needs should be deployed.

---

[2] These are FACA committees are in principle open to the entire community for input (e.g. one can call into the meetings and hear/watch the presentations on WebEx and make comments). However, in practice there are limitations that prevent some groups of people from participating including the times of the meetings and knowledge about the events.
[3] This example refers to the National Academies 2015 Committee on a Strategy to Optimize the U.S. Optical and Infrared System in the Era of the Large Synoptic Survey Telescope (LSST).
[4] We noted here is that the committee's chair was from a small teaching college, however, we point out that the main role of the chair is to build consensus rather than advocating on behalf of particular positions.

These kinds of demographic barriers (or exclusions) to advisory input from a more full cross section of the community can be mitigated, eliminated, or at least taken into account in a final report through a better understanding of who is (and is able) to contribute perspectives to an advisory group collecting community input.  By collecting and reviewing demographic information from members of the community who provide input, unintentional omissions from sectors of the community would become more clear. If done in a timely way (at or before the committee's deliberations phase), this would allow for the opportunity to expose biases or exclusions of significant sectors of the community. A committee or panel  better informed by the demographics of input contributions would then be able to decide how to mitigate or acknowledge omissions in their final report.  As with scientific research papers, this mitigation might include collecting additional data by opening a broadened collection of input (e.g., soliciting more, and perhaps targeted, white papers), directly inviting expertise testimony specifically from absent sectors of the community,  or simply acknowledging the scope of the current input along with any significant omissions. Publishing and acknowledging this demographic information along with the final report would add weight to the advisory report as being informed by the full community.

In addition to identifying exclusions or overrepresentation in data collection, there are several other reasons to consider collecting the demographic information of those who give input to advisory committees[6].  Some of these reasons are:

---

[5] We do note here that additional public input was solicited via in person invitations to speak at open committee meetings.  However, these open meetings were held only twice at locations on the east and west coasts. Thus making wide collection of  input through these means prohibitive and exclusive.
[6] Also see ASTRO2020 Decadal Survey APC paper by David W. Hogg, et al.,' A better consensus: Changes to the Decadal process itself."

Providing transparency so the community understands the process by which advice on limited resource distribution is offered; Strengthening the process by prompting buy-in to the review as a reflection of the community's recommendations and prioritization; Assuring diversity by understanding who is (and is not) taking part in the process; Assessing inclusion by providing an opportunity to understand what areas of the community are not being reached or do not feel that they have a stake in the process; Informing the committee by helping to identifying blind spots or areas of missed information.

The example described above illustrates a bias and imbalance in perspective due to institutional affiliation, however there are other unintended biases in expertise that could be introduced due to the necessarily finite size of an advisory committee or panel that come from other identity affiliations. Research on the topic of diversity in teams demonstrates that cognitive diversity is influenced by identity diversity. That is people of diverse identities bring different heuristics to thinking about and solving complex problems (Page, S., 2017). Therefore, we suggest that the following demographic information be solicited from input contributors. This information could be solicited at the time of white paper submission or in demographic surveys after submission if contact information is collected.

      a. Institutional Affiliation (with institution description and location)[7]
      b. Gender Identity, expression and/or biological sex
      c. Career phase and work status (i.e., permanent, temporary, soft money positions)
      d. Main scientific areas of expertise (using sci. keywords)
      e. Ethnic/racial identities
      f. Intersectional /multicultural identities and abilities

The appendix below gives examples of how this data might be collected, i.e., specific questions to be asked. These are suggestions, but other demographic parameters might be more (or less) relevant for collection depending on the committee and its charter. With a better understanding of where contributed perspectives have come from, committees would be able to identify any blindspots and/or weaknesses in the information gathered, as well as significant areas of overrepresentation. Therefore, to take advantage of the informative nature of this demographic data collection, it is critical that information be gathered and reviewed BEFORE the study process comes to an end.

---

[7] The 'institution description' would, for example, specify if the researchers affiliation is a small teaching college, a large state university, or other type that might suggest expertise significant to doing research at that type of institution. Descriptions could be specified via specified standard selections.

In addition to the collected data being relevant to the committee for immediate use, it would also be informational for ensuring that future reviews engage the broader community. Since an imbalance in engagement may simply be caused by inadequacies in communication or information about contributing, it is essential to collect and review demographic data to better the process.

We advocate here that advisory committees that collect outside information be able to review the demographics of submitted input and decide how they might want to use this additional knowledge in their final report. We further ***recommend that the results of this demographic information be widely available alongside the committee's final report to provide transparency of the process.***

**Conclusion:**
Peer review of scientific research is currently the best method we have for assessing and providing scientifically motivated advice on how resources should be allocated to advance the field of astronomy and astrophysics. While currently there are standards in place to help ensure expertise and balance, there may be weaknesses and blindspots that lead to bias and misrepresentation of community input on resource wants and needs. These biases can then trickle up into the final report resulting in incomplete recommendations and priority setting. However, one relatively easy and cost effective way to mitigate or eliminate these concerns is by collecting the demographic data of those individuals who contribute to the information gathering phase AND reviewing those data with the committee/panel members in a timely way. The compiled results can inform the committee's process toward a final report. This would allow the committee members to decide if any further action should be taken to provide community input opportunities. The public release of this demographic data would also provide the larger community with a better understanding of how the advisory process is done and with incentive to contribute to any future open calls for contribution.

**Appendix:**

This appendix gives suggestions for the wording of demographic survey questions. The structure of these questions has been vetted and used by Rachel Ivie, director of the AIP's Statistical Research Center.  The questions match terminology used in the AAS Workforce Survey, which has been undertaken by AAS's Demographics Committee. See https://aas.org/comms/demographics-committee for additional details and the survey results.

**Sample Survey Questions:**

•***Check here if you have already submitted information for this survey.*** *Put at beginning of survey and no need to answer any questions if already submitted survey info.*

•***What is your affiliation?***

•***Are you at a:*** *university, college, national  institute/observatory, private institute, Company/industry, other specify*

•***If at a university or college, approximately how many astronomy/astrophysics faculty does your department have?***   *<5, 5-10, 10-15, 16 or more*

•***What is the highest degree you have obtained?*** *No college or university, Bachelor, Master, Doctorate, Other (please specify)*

•***In what year did you earn your highest degree?*** *(select from list)*

•***What was your employment status on DATE?***   *(use submission deadline?)*

 *Tenure-track, Tenured, Postdoctoral, Soft money,  Long term (but not-tenured) position, Other (please specify)*

•***What are your primary areas of scientific interest?***  *(Indicate all that apply , list )*

*(Solar systems / Planetary science, Heliophysics,)Exoplanets, Astrobiology, Star formation & evolution, Interstellar medium,Galacticstructure & stellar populations, Time Domain astro, Galaxy formation & evolution, Active galactic nuclei, Clusters of galaxies/large-scale structure,Cosmology, Astronomy education, Other, please specify*

•***What is your gender identity?*** *Woman, man, Another identity, prefer not to  respond*

•***Do you identify as transgender?*** *Yes, no, prefer not to  respond.*

•***Do you identify as:*** *Heterosexual, Gay or lesbian, Bisexual,Other, Prefer not to  respond*

•***Please indicate which of the following apply to you: (Please check all that apply)***

    *I am deaf or hard-of-hearing*

*I have difficulty seeing even when wearing glasses*

*I have serious difficulty standing, walking, or climbing stairs*

*I have a cognitive or learning disability*

*I have a mental illness*

*I am neuro-atypical*

*I have an autoimmune or pain disorder, or other chronic condition*

*I have disabling allergies, asthma, or other environmental sensitivities*

*Other disability: specify*

*None of the above*

*Prefer not to respond*